# Director switching dynamics of ferromagnetic nematic liquid crystals

Nerea Sebastián,*[a] Natan Osterman,[b] Darja Lisjak,[a] Martin Čopič,[b] and Alenka Mertelj[a]

[1] J. Stefan Institute, P.O.B 3000, SI-1000 Ljubljana, Slovenia

[2] Department of Physics, Faculty of Mathematics and Physics, University of Ljubljana, Jadranska cesta 19, SI-1000 Ljubljana, Slovenia

Corresponding author: nerea.sebastian@ijs.si

Successful realization of ferromagnetic nematic liquid crystals has opened up the possibility to experimentally study a completely new set of fundamental physical phenomena. In this contribution we present a detailed investigation of some aspects of the static response and the complex dynamics of ferromagnetic liquid crystals under the application of an external magnetic field. Experimental results are then compared with a macroscopic model. Dynamics of the director were measured by optical methods and analyzed in terms of a theoretical macroscopic model. A dissipative cross-coupling coefficient describing the dynamic coupling between the two system order parameters, the magnetization and the nematic director, is needed to explain the results. In this contribution we examine the dependency of this coefficient on material parameters and the saturation magnetization and the liquid crystal host. Despite the complexity of the system, the theoretical description allows for a proper interpretation of the results and is connected to several microscopic aspects of the colloidal suspension.

## Introduction

The processability of soft or even fluid ferromagnetic/ferroelectric organic materials makes them of great interest for further technological development. Among them, liquid crystals (LC) are especially attractive candidates due to their unique combination of long-range order and fluidity. Nematic liquid crystals (NLC) are the simplest of liquid crystal phases, in which on average the molecules orient along a preferred direction corresponding to the phase director **n**. NLCs are birefringent materials with their optical axis in the direction of **n**. The possibility to control the optical response of NLC at fast switching rates by changing the direction of **n** with small external electric fields constitutes the basis of their exploitation in modern LC displays and of their consolidating role in optical imaging.[1] On the contrary, due to their low magnetic susceptibility, relatively large magnetic fields are needed to actuate nematic liquid crystals, which limits their practical utilization in magnetic based applications. Overcoming such limitation unlocks a wide range of possibilities as contact electrodes are made dispensable for magnetic control and the field can be applied in any direction. It was initially proposed by Brochard and de Gennes[2] that the dispersion of magnetic nanoparticles in a NLC would lead to an increase of the magnetic sensitivity and eventually to a room temperature ferromagnetic fluid phase, where the orientational order of the NLC would induce orientational order of the colloidal particles.

Although great effort was done in this direction and a variety of materials were prepared by the inclusion of elongated or spherical ferromagnetic particles,[3–7] the different attempts yielded paramagnetic behaviour and it immediately became apparent that overcoming particle aggregation was the crucial challenge.[8–12] While the nematic mediated elastic interaction should be strong enough to counteract the magnetic interaction between the particles, the latter should be, at the same time, of such type (quadrupolar like) to favour ferromagnetic ordering. Such a balance greatly depends on the particles size, shape and the surface treatment which will determine the preferred orientation of the LC around the particles, and thus the orientation of the particles with respect to the director of the LC host. It has not been till recently that stable ferromagnetic suspensions in a nematic liquid crystal host were achieved for Sc-substituted barium hexaferrite (BaHF) or $BaFe_{12-x}Sc_xO_{19}$ magnetic nanoplatelets suspended in 5CB.[13] The platelet shape of the particles together with perpendicular anchoring (dodecylbenzenesulphonic acid surfactant - DBSA) induce a quadrupolar nematic director field around the platelets preventing the aggregation of the platelets in the director direction. In addition, magnetic interactions between the platelets seem to be such that parallel, that is ferromagnetic, ordering of the dipoles is promoted resulting in the macroscopic magnetization **M** (Figure 1).[14] The magnetization **M** and the director **n** are thus coupled through the nanoplatelets surface anchoring of the NLCs molecules. Employing the same platelet shaped nanoparticles, ferromagnetic cholesteric liquid crystals have been successfully prepared and studied.[15–17] Also, biaxial ferromagnetic liquid crystals where the direction of the magnetization and the nematic director are at an angle, have been achieved[18] and a colloidal ferromagnetic nematic phase build on $BaFe_{12-x}Sc_xO_{19}$ suspended in an isotropic solvent has been reported.[19] A thorough review of the latest achievements on ferromagnetic nematic liquid crystals (FNLC) can be found in Reference 14.

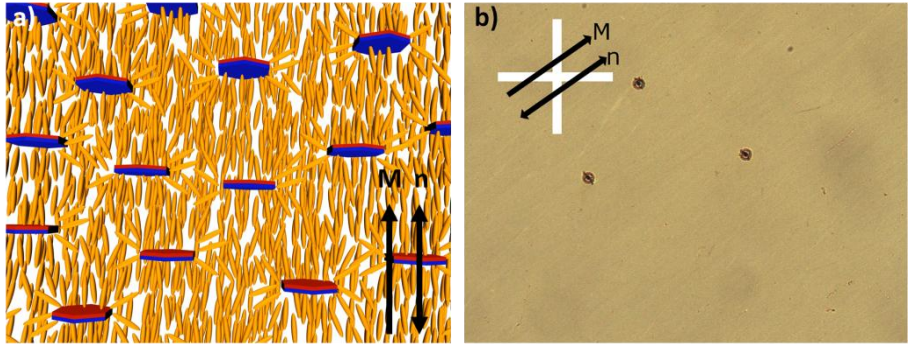

Figure 1: a) Microscopic structure of the ferromagnetic nematic liquid crystal. Platelets are represented side-on in red/blue (indicating the alignment of the magnetic moments) color and are surrounded by the nematic director field (orange schematic molecules) b) Polarizing optical microscopy image of the E7/200 nematic 20-μm sample placed between crossed polarizers, with the director at an angle of 40°. Image width corresponds to 700 μm. The spheres are cell spacers.

Due to the coupling between **M** and **n**, the reorientation of the magnetization by external magnetic fields drives the reorientation of **n**, resulting in a strong magneto-optic response.[20] Remarkably, such magneto-optic response can be used to visualize simultaneously the direction and strength of the magnetic field.[21] When the field is applied perpendicularly to the director, the reorientation occurs without threshold, in contrast to how it occurs with pure nematic liquid crystals for which a critical field needs to be exceeded. The dynamics of such switching have been recently thoroughly examined[22,23] demonstrating that a dissipative dynamic coupling between **M** and **n** should be taken into account in order to address the observed behaviour. This dynamic coupling strongly influences the switching times by making the response noticeably faster, which is essential for their exploitation in magneto-optic devices. The origin of this dissipative dynamic coupling remains unclear and it has been speculated to arise from the coupling between the director orientation and the microscopic fluid flow localized in the vicinity of the rotating magnetic platelets.[22]

The aim of this paper is to confront the experimental results with a detailed macroscopic model, in order to shed light on the origin of the static and the dynamic coupling between **M** and **n**. We examine the dependence of the dissipative dynamic cross-coupling coefficient on the magnetization/concentration value and on the nematic host flow orienting ability. For this purpose, the dynamics of the reorientation of **n** has been resolved by determining the phase difference between the ordinary and extraordinary light (Methods) for different suspensions as a response to the application of a small perpendicular magnetic field and analyzed against the theory.[22,23] This paper is organized as follows. The macroscopic description of the suspension is given in the section Theoretical background. Materials and the main results are then thoroughly detailed in the Results section. Finally, results are examined in the Discussion section.

## Theoretical Background

Macroscopically a ferromagnetic nematic phase can be described by two coupled order parameters: the nematic order parameter $\mathbf{Q} = S(\mathbf{n} \otimes \mathbf{n} - 1/3\underline{\mathbf{I}})$, where the scalar parameter $S$ accounts for the degree of ordering of the molecules along the director **n**, and the magnetization $\mathbf{M} = M_0 \mathbf{m}$, where **m** is a unit vector. Considering constant temperature, pressure, uniform concentration of particles, constant magnetic field **H** and electric displacement field **D**, the free energy density of the system can be written as[13,20]

$$f = f_0 + f_{elastic} + f_{coupling} + f_{field}, \qquad 1$$

where $f_0$ represents the free energy describing the homogeneous ferromagnetic nematic phase and only depends on $S$ and on the magnitude of **M**, i.e, is constant under the mentioned assumptions. Any director distortions implies an elastic contribution $f_{elastic}$ given by the Frank elastic energy:

$$f_{elastic} = \frac{1}{2}K_1(\nabla \cdot \mathbf{n})^2 + \frac{1}{2}K_2(\mathbf{n} \cdot (\nabla \times \mathbf{n}))^2 + \frac{1}{2}K_3(\mathbf{n} \times (\nabla \times \mathbf{n}))^2, \qquad 2$$

with $K_1$, $K_2$ and $K_3$ being the splay, twist and bend elastic constants, respectively. The coupling between the two order parameters is given by the coupling term

$$f_{coupling} = -\frac{1}{2}\kappa \mu_0 M_0^2 (\mathbf{n} \cdot \mathbf{m})^2, \qquad 3$$

with the sign of the coupling term $\kappa$ conditioning the relative orientation of **m** with respect to **n**. The contribution of the coupling of **n** and **m** with external fields is given by

$$f_{fields} = -\mu_0 M_0 \mathbf{m} \cdot \mathbf{H} + \frac{1}{2}\mathbf{D} \cdot \mathbf{E}. \qquad 4$$

The coupling strength with **E** is determined by the dielectric anisotropy $\varepsilon_a$, being $\varepsilon_0 \mathbf{E} = \underline{\boldsymbol{\varepsilon}}^{-1}\mathbf{D}$ and the dielectric tensor $\underline{\varepsilon} = \varepsilon_\perp \mathbf{I} + \varepsilon_a \mathbf{n} \otimes \mathbf{n}$ with $\varepsilon_\perp$ the dielectric constant perpendicular to the director. It should be remarked that a complete description of the free energy density should also include the terms corresponding to the direct coupling of the director $\mathbf{n}$ to $\mathbf{H}$, an elastic term associated to the deformation of $\mathbf{M}$ and a stray field energy term. However, these terms have been estimated to be at least two orders of magnitude smaller than those described here for the values of $M_0$ of the order of those achieved in the samples under consideration and they can be reasonably neglected in our case.[14]

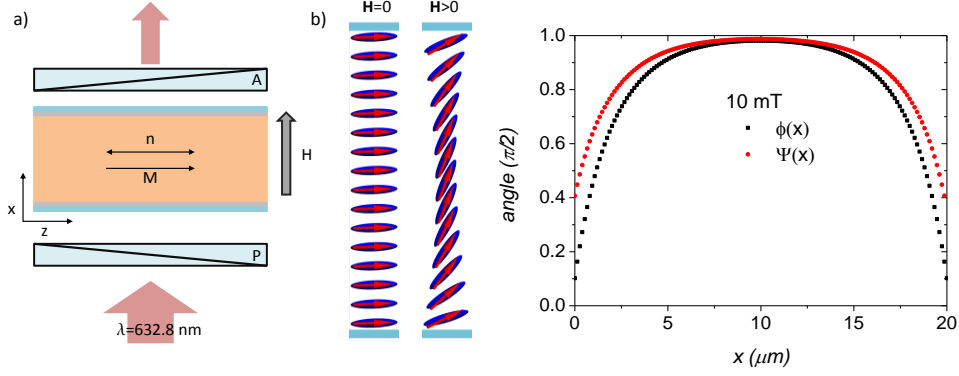

Figure 2: a) Schematic representation of the experimental setup. b) Representation of the director and magnetization distortions across the liquid crystal cell under the application of an external field perpendicular to the cell normal.

Finally the effect of the anchoring of the surfaces can be taken into account by the additional surface term[24]

$$f_{surface} = -\frac{1}{2} W (\mathbf{n} \cdot \mathbf{n}_s)^2 \qquad 5$$

with $W$ being the anchoring strength and $\mathbf{n}_s = (\sin\varphi_s, 0, \cos\varphi_s)$ the preferred direction at the surfaces given by the pretilt angle $\varphi_s$. In our experiments, commercial cells with strong planar anchoring ($W \approx 10^{-4}$ Jm$^{-2}$) and a pretilt angle between 1° and 3° were employed.

Application of a magnetic field perpendicularly to the cell surface, i.e, perpendicular to $\mathbf{m}$, will cause the reorientation of the nanoplatelets and this drags the reorientation of the director due to the coupling between $\mathbf{n}$ and $\mathbf{m}$ through the platelets surface anchoring. In recent extensive theoretical studies, it has been shown that, in order to explain the measured relaxation rates under the application of such a magnetic external field, a dissipative cross-coupling between $\mathbf{n}$ and $\mathbf{m}$ should be included in the dynamic equations, which then read as

$$\dot{\mathbf{n}} = -\frac{1}{\gamma_1} \mathbf{h}_n^\perp + \underline{\chi}^D \mathbf{h}_m^\perp$$
$$\dot{\mathbf{m}} = \frac{M_0^2}{\gamma_{mag}} \mathbf{h}_m^\perp + \underline{\chi}^{D,T} \mathbf{h}_n^\perp. \qquad 6$$

where $\gamma_1$ is the rotational viscosity, $\gamma_{mag}$ is the magnetic rotational viscosity (analogous to $\gamma_1$) and $\underline{\chi}^D$ depends on the scalar cross-coupling coefficient $\chi_2^D$.[22,23] Here the components of the molecular fields $\mathbf{h}_n^\perp$ and $\mathbf{h}_m^\perp$ correspond to the projection into the plane perpendicular to the director and magnetization of the nematic, respectively. The molecular fields are calculated as

$$\mathbf{h}_n = \frac{\partial f}{\partial \mathbf{n}} - \frac{\partial}{\partial x} \frac{\partial f}{\partial (\partial \mathbf{n}/\partial x)}$$
$$\mathbf{h}_m = \frac{\partial f}{\partial \mathbf{m}}, \qquad 7$$

and due to the constraints $\mathbf{n}^2 = 1$ and $\mathbf{m}^2 = 1$, only $\mathbf{h}_n^\perp$ and $\mathbf{h}_m^\perp$ enter dynamic equations 6.

The behaviour described is quite complex and the equations depend on many material parameters which results in complicated analysis of experimental results. Conveniently, it has also been shown that the initial dynamics of the normalized phase difference between the ordinary and extraordinary light (see Methods) $\phi_n(H) = 1 - \phi(H)/\phi(0)$ simplifies to[22,23]

$$\phi_n(H,t) \approx \frac{n_{e0}(n_{e0}+n_o)}{2n_o^2}\left[\left(\chi_2^D M_0 \mu_0 H\right)^2 t^2 + 2\varphi_s \chi_2^D M_0 \mu_0 H t\right]$$
$$\equiv k_{ini}^2 t^2 + pt \qquad 8$$

where $n_{e0}$ and $n_{o0}$ are the extraordinary and ordinary refractive indices. Such expression reveals that at short times, when the distortions of $\mathbf{M}$ and $\mathbf{n}$ are small, dynamics is directly dictated by the dissipative cross-coupling coefficient. Conveniently, by means of this expression, the dissipative cross-coupling coefficient $\chi_2^D$ can be determined from the field dependence of the $k_{ini}$ coefficient, with the sole requirement of knowing the sample magnetization $M_0$ and indices of refraction.

We measured the magnetization and phase difference according to the geometry depicted in Figure 2. According to this geometry it can be assumed that the physical properties depend only on the coordinate along the cell thickness. Thus, choosing the x axis of the coordinate system perpendicular to the cell surface, i.e, in the direction of the external field and the z axis along the initial director, the director and magnetization can be written as $\mathbf{n}=(\sin\varphi(x,t),0,\cos\varphi(x,t))$ and $\mathbf{M}/M_0=(\sin\psi(x,t),0,\cos\psi(x,t))$. The measured magnetization and the phase difference can then be expressed as

$$M_x = \frac{M_0}{d}\int_0^d \sin\psi(x,t)\,dx \qquad 9$$

and

$$\phi = \frac{2\pi}{\lambda}\int_0^d \left(\frac{n_o n_e}{\sqrt{n_e^2\cos^2\varphi(x,t)+n_o^2\sin^2\varphi(x,t)}}-n_0\right)dx, \qquad 10$$

where the time dependence of $\varphi(x,t)$ and $\psi(x,t)$ results from the solution of equations 6.

## Results

### Sample characterization

Sc-substituted barium hexaferrite nanoplatelets ($BaSc_xFe_{12-x}O_{19}$) were synthesized hydrothermally.[25] The nanoplatelets thickness is around 5 nm and the diameter distribution is approximately log-normal, with a mean of 60 nm and a standard deviation of 20 nm. As described in References 14 and 20 ferromagnetic nematic suspensions have been prepared by mixing stable suspension of the nanoplatelets coated with dodecylbenzenesulphonic acid (DBSA) in t-butanol and the different nematic hosts (E7, 7CB and 5CB, see Table 1) in the isotropic phase. After the evaporation of the solvent, the remaining stable suspensions were quenched from the isotropic to the nematic phase. The DBSA surfactant promotes perpendicular orientation of the nematic director at the particles surface. In this way, the particles orient themselves with their short axis, i.e., their magnetic moment, parallel to $\mathbf{n}$.[13] Finally, the nematic suspension was centrifuged to remove any aggregates and 20 μm planar LC cells (Instec Inc.) with ITO and rubbed polyimide surfaces were filled. Cells where filled under a magnetic field of 8 mT along the rubbing direction in order to easily obtain magnetic monodomain samples with the magnetization $\mathbf{M}$ parallel to $\mathbf{n}$. (Figure 1)

In order to characterize the suspensions, we studied the x-component of the magnetization as a response to the application of small external fields perpendicular to the direction of the undistorted $\mathbf{n}$ and $\mathbf{M}$. In the absence of magnetic field $M_x$ is negligible, arising from the small pretilt angle of the surface treated cells. Upon application of the field, $M_x$ initially increases linearly and saturates at fields no much larger than 10 mT (Figure 3.a). The volume concentration of the platelets (c) can be estimated from the saturation value of the magnetization $M_0$ (Table 1).

### Static coupling parameter

The unique range of properties exhibited by the FNLCs is a direct consequence of the coupling between the director and the magnetization through the anchoring of the NLC molecules to the platelets' surface. When applying a magnetic field perpendicularly to the initial $\mathbf{n}$ and $\mathbf{M}$, a torque is exerted on $\mathbf{M}$ causing it to reorient in the field direction. The coupling between $\mathbf{n}$ and $\mathbf{M}$ causes the reorientation of $\mathbf{n}$ in the same direction, balanced by the torque at the cell surfaces due to the cells anchoring treatment. The reorientation takes place with no threshold, which is a notable difference from typical Frederiks transition of LCs. The change of optical properties resulting from the reorientation of $\mathbf{n}$ can be used to determine the value of the static coupling parameter $\kappa$. Similarly, the constrains on $\mathbf{n}$ imposed by the cell surfaces would impact the shape of the magnetization curves. Thus, in order to obtain an estimate of $\kappa$ the dependence of both, the magnetization $M_x$ and the normalized phase difference (Figure 3), on the external magnetic can be fitted by numerical minimization of the free energy (detailed description of the fitting procedure can be found in reference 20). The material and cell parameters values $\{K_1, K_3, \Delta n, W, \varphi_s\}$ were determined in advance from the measurements of the electrical Frederiks transition (see Methods) and fixed in order to limit the number of fitting parameters and their effect on the value of $\kappa$. The measured optical response and magnetization curves are plotted together with the corresponding theoretical fits (solid lines) in Figure 3. Following the same procedure the static coupling parameter for a variety of 5CB and 7CB samples has also been determined, and the values are given in Table 1.

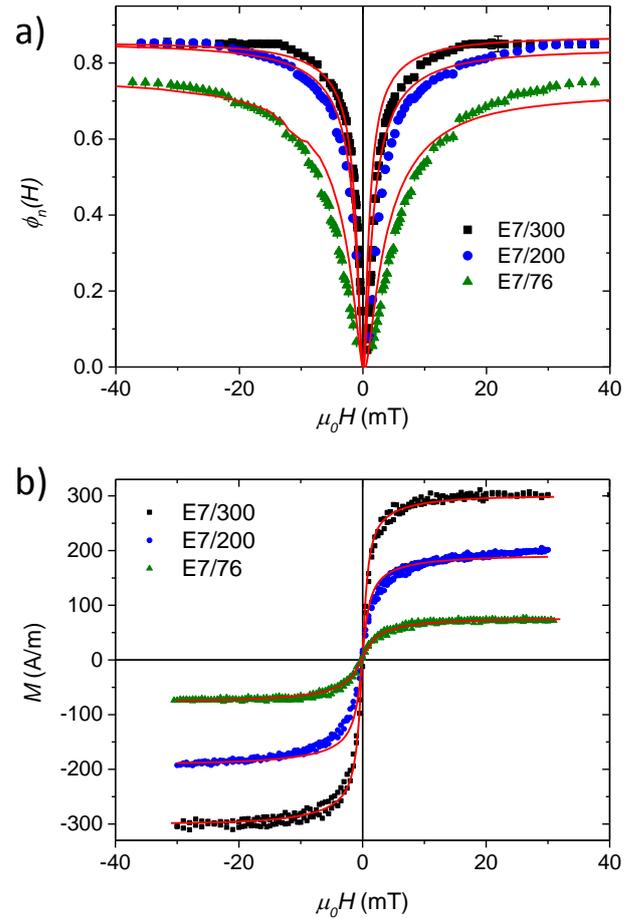

Figure 3: a) Magneto-optic effect in a perpendicular field shown by the normalized phase difference vs. external magnetic field. b) Magnetization curve measured for transverse field. Lines in both graphs are fits obtained by energy minimization (see text).

Table 1. Obtained material sample parameters.

| Sample | $M_0$ (A/m) | c | $\kappa$ | $\chi_2^D$ (Pa s)$^{-1}$ | $\gamma_{mag}$ mPa s |
|---|---|---|---|---|---|
| E7/76 | 76 | 0.4·10⁻³ | 210±10 | 4.2±0.7 | 0.23 |
| E7/200 | 200 | 1·10⁻³ | 70±10 | 4.2±0.7 | 0.30 |
| E7/300 | 300 | 2·10⁻³ | 35±10 | 4.2±0.7 | 0.29 |
| 5CB/90 | 90 | 0.45·10⁻³ | 160±10 | 11.5±0.3 | 0.085 |
| 7CB/165 | 165 | 0.8·10⁻³ | 88±5 | 6.7±0.2 | 0.145 |

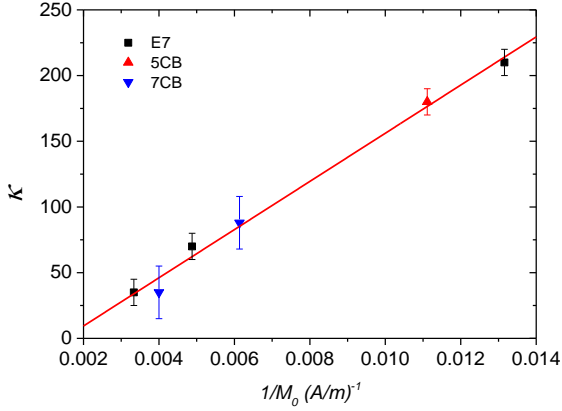

Figure 4: Value of the static coupling parameter $\kappa$ vs the inverse of the saturation magnetization $M_0$. The linear behaviour is in agreement with Eq 11.

On the microscopic level, considering independent particles, it was proposed that the coupling energy density can be related to the total surface energy of an individual platelet given by $2S_0 F_{plat} = -S_0 W_{plat} (\mathbf{n} \cdot \mathbf{n}_{plat})^2$ where $S_0$ corresponds to the platelet surface, $W_{plat}$ is the anchoring strength on the particle surface due to the surfactant and $\mathbf{n}_{plat}$ is the preferred direction of the director at the surface of the particle (parallel to $\mathbf{n}$ in the present case). Equating equation 3 with the total surface energy of all the platelets given by the volume fraction ($c$), we can write

$$\frac{1}{2}\kappa\mu_0 M_0^2 \approx \frac{cW_{plat}}{d_{plat}}. \qquad 11$$

Taking into account that $M_0$ is directly proportional to the volume fraction, a dependency of $\kappa \propto 1/M_0$ is expected from expression 11, which is in very good agreement with our results (Figure 4).

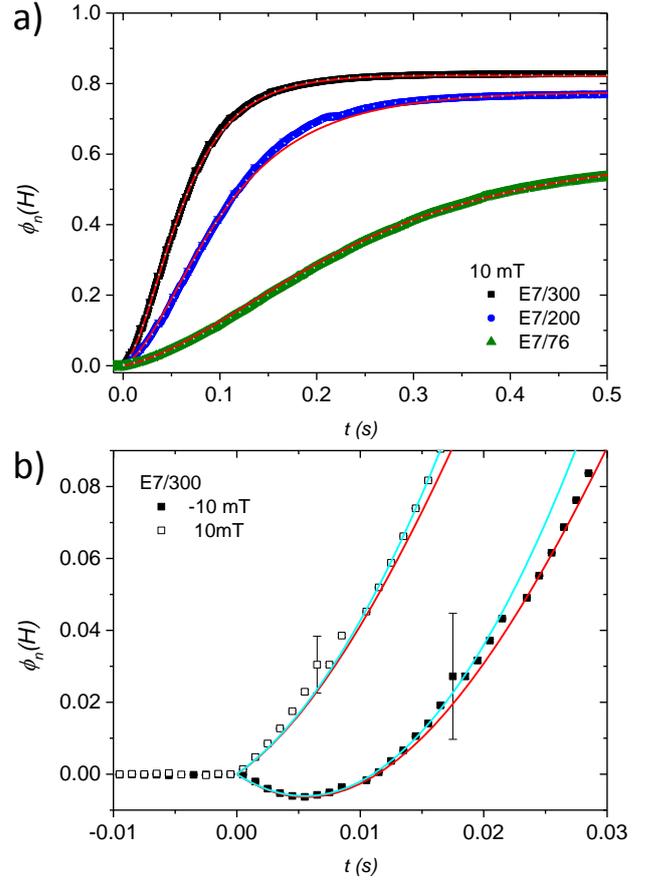

Figure 5. a) Time evolution of the measured normalized phase difference for the three E7 samples when applying a 10 mT magnetic field. Dashed lines correspond to the fits according to equation 12. Red solid lines correspond with theoretical dynamic model calculated from equations 6. Theoretical and analytical fit adequately overlap. b) Initial dynamics of the measured normalized phase difference for the E7/300 sample at 10 and -10 mT. Solid red lines correspond with theoretical dynamic model calculated from equations 6. Solid blue lines correspond to the linear-quadratic fit according to equation 8. Experimental error bars are derived from the $\arcsin(I/I_0)$ (see Methods) which diverges at the position of the singular points.

From the obtained linear behaviour we can estimate the value of the platelets anchoring strength to be $W_{pl} \approx 1.2 \cdot 10^{-5}$ J m$^{-2}$. Such value is typical for an intermediate ($K/W \sim 1$ μm) strength, neither weak nor strong. However, it should be noted that this value is obtained neglecting the polydispersity of the platelets and considering them independent and so, it should be just considered as an estimate.

### Dissipative dynamic cross-coupling parameter

The time evolution of $\phi_n(H,t)$ upon switching on the external magnetic field was recorded systematically for several strengths of the field for the different E7 samples (Figure 5). In first place the initial dynamics behaviour was analysed and fitted to the linear-quadratic expression 8 (Figure 5.b). In the case of negative fields, a clear minimum is detected. It arises from the cell´s pretilt, which determines the initial director field in a way that, when reorienting on the field, the director field crosses the zero angle state at the onset of the reorientation. For positive pretilt angles, such minimum occurs for negative fields, while for negative pretilt angles it is detected at positive fields. Thus, the relative sign depends only on the selection of the sign of the coordinate axis. The curves with the minimum (negative fields, in our case) allow for a more accurate determination of the validity range of the initial dynamics linear expression (equation 8) at each magnetic field, which can be then extrapolated also for the positive fields fit. The resulting values for the initial switching rate are shown in Figure 6. The initial rate $k_{ini}$ is predicted to depend linearly on the field strength by $\chi_2^D M_0 \sqrt{r_0}$, where $r_0$ is the prefactor in equation 8, which depends on the refraction indices. For the ordinary refractive index the value of 1.52 was assumed which is in agreement with literature.[26] Then, the extraordinary index is written as $\Delta n + n_o$, where the birefingence value is taken from our measurements. The normalization of $k_{ini}$ by $M_0\sqrt{r_0}$ (inset in Figure 6) shows that the value of the dissipative cross-coupling coefficient between the director and the magnetization $\chi_2^D$ is independent of $M_0$, i.e., from the concentration of nanoparticles in the FNLC. Significantly, this implies that, for a given liquid crystal host, the value of $\chi_2^D$ seems to be given by an intrinsic parameter of the host. From the linear fit of the data confined in Figure 6 we can determine that the value of $\chi_2^D$ to be (4.2±0.7) (Pa s)$^{-1}$ in the case of the liquid crystal host E7.

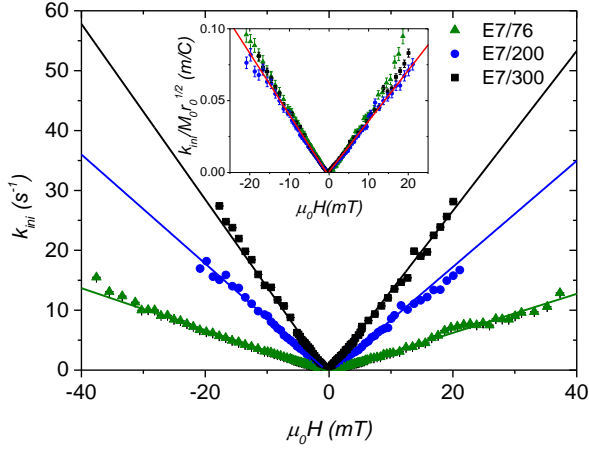

Figure 6. Magnetic field dependence of the initial dynamics coefficient $k_{ini}$ (equation 8). The dynamic cross-coupling coefficient $\chi_2^D$ can be obtained from the straight line fits. The inset shows the collapse of the three sets of data under the normalization of $k_{ini}$ by $M_0\sqrt{r_0}$.

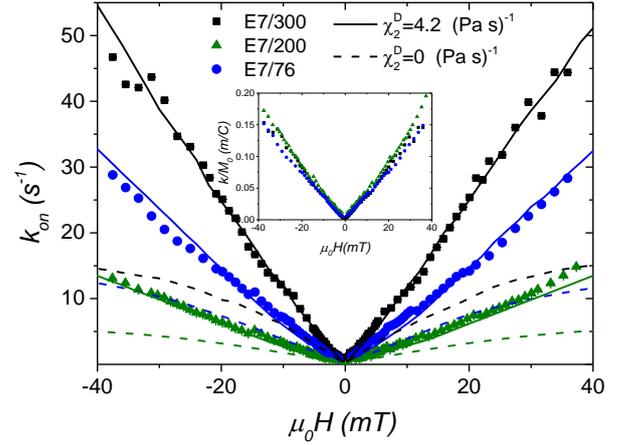

Figure 7. Magnetic field dependence of the overall switching rate extracted from the fits of the data to equation 12. Solid lines correspond to the calculated rate obtaining by solving equations 6 for a dissipative cross-coupling coefficient $\chi_2^D = 4.2$ (Pa s)$^{-1}$. Dashed lines show that theoretical rate calculated without taking into account the dynamic cross-coupling, strongly deviates from the obtained experimental results. Inset shows the unique field dependency of the switching rates normalized by the magnetization for the three E7 samples.

The characteristic switching rate $k_{on}$ describing the full dynamics was obtained by fitting each data set to a squared sigmoidal function[22]

$$f(t) = c'\left(1 - \frac{1+c}{1+c\exp(-2k_{on}t)}\right)^2. \qquad 12$$

An example of such fits is given in Figure 5.a (dashed lines) for all the three samples upon application of 10 mT. It can be observed in Figure 7 that the obtained field strength dependency of the characteristic switching rate for the three E7 samples is linear and scales with the absolute value of the magnetization, as shown in the inset. In order to compare the measured rates with those that can be calculated by solving the equations 6, two additional dynamic parameters need to be known: the director rotational viscosity $\gamma_1$ and the magnetic rotational viscosity $\gamma_{mag}$. In order to limit the number of parameters to be extracted from the theory, the rotational viscosity $\gamma_1$ was independently obtained from dynamic light scattering (DLS) measurements (see Methods) to be 155 mPa·s. The data and the model were then confronted by varying the value of $\gamma_{mag}$ and taking into account the restrictions imposed by the condition $\chi_2^D < \sqrt{1/\gamma_1\gamma_{mag}}$ for entropy production.[22] As can be seen in Figure 5 (solid lines) the theoretically obtained time evolution of the normalized phase difference accurately describes the measured data in all the cases. The obtained values of $\gamma_{mag}$ were around 290 mPa·s, 300 mPa·s and 230 mPa·s for the E7/300, E7/200 and E7/76 respectively, which are the same within the experimental error. From the model, the linear field strength dependency is also recovered, which is in reasonably good agreement with the measured data (solid lines in Figure 7). To quantify the effect of $\chi_2^D$ on the dynamics, we calculated the expected switching rate for $\chi_2^D = 0$ (dashed lines in Figure 7). It is evident from the comparison, that the measured switching dynamics for FNLC are faster than expected from a simplified picture of the dynamics and that the dissipative cross-coupling coefficient plays an important role which cannot be disregarded.

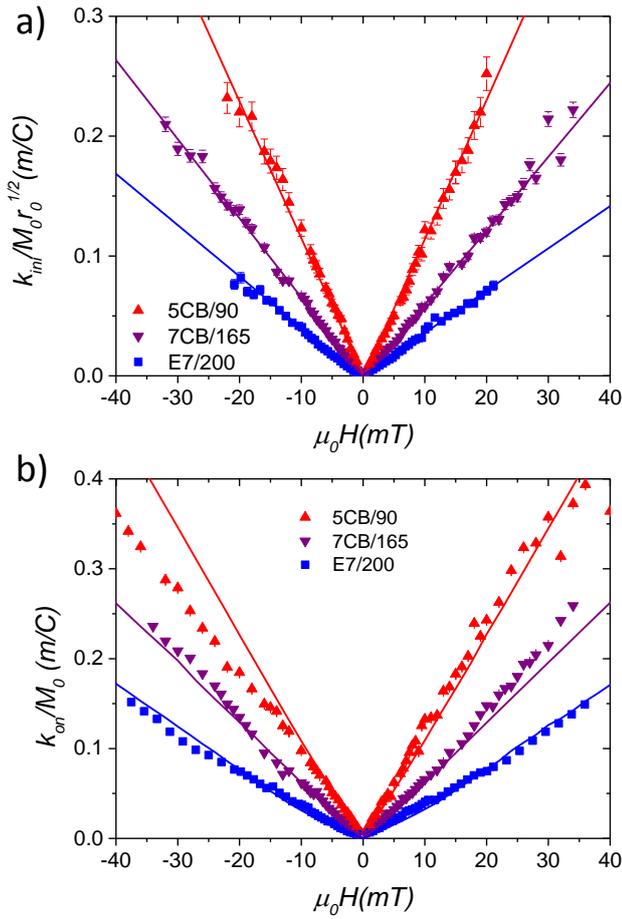

Figure 8. a) Magnetic field dependence of the normalized initial dynamics coefficient $k_{ini}/M_0 r_0^{1/2}$ (equation 8 ) for FNLC with different liquid crystal hosts. b) Magnetic field dependence of the overall switching rate $k_{on}$ normalized by the magnetization extracted from the fits of the data to equation 12. Solid lines correspond to the calculated rate obtaining by solving equations 6 for a dissipative cross-coupling coefficient $\chi_2^D = 4.2$ (Pa s)$^{-1}$, $\chi_2^D = 6.7$(Pa s)$^{-1}$ and $\chi_2^D = 11.5$ (Pa s)$^{-1}$ for E7/200, 7CB/165 and 5CB/90 respectively.

Now, the question of how does $\chi_2^D$ depend on the liquid crystal host arises. With the aim of answering this question we prepared ferromagnetic suspensions with the 7CB nematic liquid crystal host, to be compared with the results from E7 and 5CB.[22] Both, 5CB and 7CB, are single component liquid crystals showing nematic phase at room temperature, while E7 is a four component mixture with the firsts as main components. Interestingly, according to reports, 5CB is a strong flow-orienting nematic while the 7CBs flow orienting ability is significantly lower[27,28] or is even close to tumbling and E7 lies in between.[29] As in the case of E7, rotational viscosities have been measured by means of DLS (see Methods) and the obtained values are 140 mPa·s and 75 mPa·s for 7CB and 5CB, respectively. Three samples have been compared, E7/200, 5CB/90 and 7CB/165. Although the magnetization of the 5CB sample greatly differs from the other two, it has been just shown that such differences do not influence the value of $\chi_2^D$ deduced from the initial dynamics. We performed the initial dynamic analysis for the 5CB and 7CB samples. The comparison of the normalized initial switching rate (Figure 8) shows that the value of the dissipative cross-coupling coefficient drastically changes for the three liquid crystals. As in the previous case, assuming $n_o = 1.52$ in both cases,[30] we obtain the values (11.5±0.3) (Pa s)$^{-1}$ and (6.7±0.2) (Pa s)$^{-1}$ for the 5CB and 7CB respectively. They can be compared to that of (4.2±0.7) (Pa s)$^{-1}$ obtained in the case of E7. Finally, in the same way as for the E7 samples, the analysis of the full dynamics has been performed. We have previously seen that the switching rate for a given LC host scales with the saturation magnetization, and so, in order to compare Figure 9 shows the theoretical calculated magnetic field dependencies of the switching rates normalized by $M_0$ and the experimentally determined ones. Once again the theoretically calculated rates are in very good correspondence with the experimental values. Finally, the values of the magnetic viscosity derived from the fits were found to be similar to those of rotational viscosity, taking values around 140 mPa·s and 70 mPa·s for 7CB and 5CB respectively.

## Discussion

In the present work we have analysed different aspects of ferromagnetic nematic liquid crystals against a macroscopic description of this complex systems, based on the macroscopic magnetization **M** and nematic director **n** parameters. A direct coupling term whose strength is defined by $\kappa$ between both parameters constitutes the key term in the free energy of the system. As has been described such term drives unique magneto-optic behaviour of the FNLC. It is noteworthy that such macroscopic static coupling parameter can be directly related to microscopic parameters of the system as the concentration of particles, their magnetic moment, size and shape together with their surface anchoring properties. Despite the polydispersity of the platelets, a linear trend is observed between the coupling parameter and the inverse of the saturation magnetization value, from which a reasonable value of the platelet surface anchoring strength could be deduced. This represents a nice example of a detailed correspondence between macroscopic and microscopic description of a complex fluid. The agreement of the data and the theory predictions is very good, where the deviation observed at low field for the normalize phase difference could be attributed to the polydispersity of the platelets, which could lead to an individual platelet rotation at low torques. Nevertheless, the simultaneous correlation of the theory with two sets of independent data, one related to the director field while the other reflects the magnetization director field, reinforces the validity of the results.

In terms of the hydrodynamics of FNLC, one should keep in mind that not only **n** and **M** are statically coupled, but they are coupled to the flow and dynamically to each other. In this context, we studied the dynamics of the director reorientation under the application of an external magnetic field. In particular, we focused on the effect and underlying source of a dissipative cross-coupling coefficient $\chi_2^D$, which is essential to describe the observed rates. The analysis of the dynamics for samples with the same LC host but different particles concentration proves that the value of the dissipative cross-coupling coefficient is independent of the sample's magnetization, i.e. of the macroscopic picture, but is a constitutive parameter of the host. The determination of different values when varying the LC host unambiguously supports that idea, i.e. that the strength of $\chi_2^D$ depends on molecular parameters given by the liquid crystal. Given the Leslie viscosities $\alpha_2$ and $\alpha_3$, the tumbling parameter defined as $\lambda = (1 + \alpha_3/\alpha_2)/(1 - \alpha_3/\alpha_2)$ determines that for $\lambda > 1$ the material is a flow-aligning nematic while it is tumbling for $\lambda < 1$. For the three hosts considered here values of $\lambda$ ordered 5CB>E7>7CB have been reported,[27–29] with 7CB being close to tumbling. However, this trend is not directly reflected in the observed values of the dynamic cross-coupling coefficient, which indicates that the origin of this dynamic parameter is more intricate than directly related to the tumbling parameter. Additional question arises whether E7, being a mixture whose main components are 5CB and 7CB, should follow the trend or not, as the different components could show different affinity for the platelets anchoring and thus, microsegregate in the surrounding. Unfortunately the results presented here do not allow for a conclusive answer to this question. In any case, with the data available, both, the dynamic cross-coupling coefficient and the switching rate, are directly related to the rotational viscosity. The lower the value of $\gamma_1$ the faster the dynamics and the larger the value of $\chi_2^D$. With all of this, it would be interesting to perform similar analysis at different temperatures in the nematic phase of a well know tumbling nematic material whose rotational viscosity increases on approaching the transition to the smectic phase, like 8CB, in order to be able to give a further description. Furthermore, the observed scaling of the full switching dynamics with the magnetization saturation value should be noted at this point. All together, the selection of an appropriate LC host together with achieved larger magnetizations could lead to much faster switching response for FNLCs. Finally, we would like to emphasize that, according to the experimental results presented here, while the static coupling parameter is not very sensitive to the choice of the liquid crystal host, dynamic response is greatly determined by it.

Regarding the magnetic rotational viscosity $\gamma_{mag}$, analogous to the rotational viscosity $\gamma_1$, in the case of E7, values vary from 300 mPa·s to 230 mPa·s, around two times that measured for $\gamma_1$ (150 mPa·s). It can be shown[31] that in the case of independent disks the ratio between magnetic and director rotational viscosities is expected to be around 0.032. The big discrepancy arises from the wrong assumption that the platelets are independent. For large enough concentrations, particles are closer to each other and the flow around a rotating disk counteracts the rotation of the neighbouring ones. Thus it seems reasonable to assume that the larger the concentration the larger the effective magnetic viscosity, which agrees with the slightly lower value obtained for the E7/76 sample. However, our results seem to point that there is a limiting value for the magnetic viscosity, which should be influenced by other parameters in the system. One should keep in mind that the FLNC is a liquid system in which the platelets are free to move, and so, any dynamic parameter or spatial rearrangement of the platelets can play a role. On the other hand, it is interesting to note that the ratios $\gamma_{mag}/\gamma_1$ determined in the case of 5CB and 7CB are around 1.

Finally, it should be remarked that the theory used in this paper[13,22,23] constitutes a exceptional example of a successful macroscopic description of a complex system. Not only it achieves an accurate interpretation of all the unique static effects found in FNLC, like magneto-optic or converse magneto-electric effects,[20] but also reliably accounts for the effect of the dynamic coupling between **n** and **M** on the system dynamics. This coupling makes the hydrodynamic equations very complex, which complicates the characterization and description of the system. In this context, the study of the switching dynamics turns out to be a powerful tool to disembroil and simplify the problem. While from the static measurements parameters such as elastic constants and static coupling parameter can be determined, from the switching dynamics the two main characteristic ferromagnetic parameters can be obtained, i.e., the dissipative cross-coupling coefficient and the ferromagnetic effective viscosity. With all this parameters one is then in the position to experimentally study and subsequently use the theory to interpret other aspects of the dynamic behaviour of FNLCs that have been theoretically predicted, such as the fundamental hydrodynamic excitations of **n** and **M** and their dependency on external magnetic fields,[20,23,31] hydrodynamic pattern instabilities, the effect of flow on the FNLC dynamics[32] or their magneto-rheological behaviour.[33] All this phenomena are far from being understood, however comprehension of FNLC hydrodynamics and its combined magneto- and electrorheological properties is a prerequisite for its prospective applications.

## Experimental Methods

### Magnetization measurements

The magnetization curves of all the samples were obtained employing a vibrating-sample magnetometer (LakeShore 7400 Series VSM). The LC cell was rotated with respect to the applied magnetic field in order to obtain the magnetization curve for parallel and transverse fields with respect to the rubbing direction.

### Measurements of the phase difference

Changes of the director orientation were determined by the measurement of the phase difference between the ordinary and extraordinary transmitted light using polarizing microscopy. In this way, the measured area is also imaged and so, any contribution from imperfections can be avoided and the uniformity of the response can be monitored during the measurement. The sample was placed so as the director was at 45° with respect to crossed polarizers and the magnetic field was applied perpendicularly to the cell plates. The time dependence of the intensity of the transmitted monochromatic light (obtained by using an interference filter with λ=632.8 nm to filter the halogen lamp of the microscope) was recorded with a CMOS camera (IDS Imaging UI-3370CP) at 997 fps. This method allows us to monitor during the whole experiment that the response is uniform and that aggregates are not present in the signal recording area. From the transmitted intensity $I = I_0 \sin^2(\phi(H)/2)$ the normalized phase difference $\phi_n(H) = 1 - \phi(H)/\phi(0)$ can be determined, where $\phi(0)$ corresponds to the phase difference in zero magnetic field.

**Determination of basic material parameters**

Material elastic constants were determined from the standard electric Frederiks transition measurements in 8 μm planar cells. Frederiks transition was studied both optically (see previous section) and dielectrically (Agilent E4980A). Curves where fitted analytically[34,35] and by energy minimization.[20] The latter allows for an estimation of the cell's surface anchoring strength. Rotational viscosity of the liquid crystals were determined at room temperature by means of dynamic light scattering (DLS) measurements (frequency-doubled diode-pumped ND:YAG laser (532 nm, 80 mW), an ALV APD based "pseudo" cross correlation detector, ALV-6010/160 correlator and single mode optical fiber with a GRIN lens was used to collect the scattered light within one coherence area). By choosing the appropriate geometry, pure modes can be measured with relaxation rates $1/\tau_i = K_i q^2 / \eta_i$.[24] In the case of twist geometry $\eta_2$ is directly the rotational viscosity $\gamma_1$. In the case of splay geometry the viscosity $\eta_1 = \gamma_1 - \alpha_3^2/\eta_b$ is affected by the backflow, which is in the case of cyanobiphenyls and E7 in general small[27,29,36,37] and so $\eta_1 \approx \gamma_1$. The obtained values are 150 mPa·s, 140 mPa·s and 75 mPa·s for E7, 7CB and 5CB respectively.

## Conflicts of interest

There are no conflicts to declare.

## Acknowledgements

NS thanks the "European Union's Horizon 2020 research and innovation programme" for its support through the Marie Curie Individual fellowship No. 701558 (MagNem). The authors acknowledge also the financial support from the Slovenian Research Agency (MČ and AM research core funding No. P1-0192, DL research core funding No. P2-0089 and AM and DL the project No. J7-8267). We thank the CENN Nanocenter for use of the LakeShore 7400 Series VSM vibrating-sample magnetometer.

## Notes and references